\documentclass[prl,floatfix,twocolumn,showpacs,preprintnumbers,amsmath,amssymb,superscriptaddress]{revtex4}
\usepackage{graphicx,float,datetime}
\usepackage[ansinew]{inputenc}

\newcommand{\udt}[3]{#1^{#2}_{\phantom{#2}#3}}

\newdateformat{mydate}{\THEDAY\hspace{3pt}\monthname[\THEMONTH] \THEYEAR}

\begin{document}

\begin{center}
\title{Cosmological Effects on Black Hole Formation}
\date{\mydate\today}
\author{Jackson Levi Said}
\affiliation{Physics Department, University of Malta, Msida, MSD 2080, Malta}
\author{Kristian Zarb Adami}
\affiliation{Physics Department, University of Malta, Msida, MSD 2080, Malta}
\affiliation{Physics Department, University of Oxford, Oxford, OX1 3RH, United Kingdom}

\begin{abstract}
{The formation of cosmological black holes is investigated using the functional Schrödinger equation as observed by an asymptotic observer, assuming a spherical domain wall collapse process. The mass formula of the Sultana-Dyer black hole is derived using Israel's domain wall mathematical framework. This is used to examine the semi-classical and quantum nature of the collapsing domain wall in the general scale of an arbitrary scale factor, while ignoring evaporation and back reaction mass losses. Particular FLRW scale factors are then explored for black hole horizon formation times, all yielding the same over-all classical result, namely that an arbitrary amount of time is required for an asymptotic observer to register the formation of this surface.}
\end{abstract}


\maketitle

\end{center}

\section{I. Introduction}
The process and mechanism by which black holes take formation has been an enduring enigma that has eluded speculation and investigation since the first exact solution was presented by Schwarzschild \cite{p11}. We explore this formation process in light of the new frameworks developed out of quantum theory and the increasingly realistic models of domain wall collapse theory \cite{p5,p8,p9,p10}. In particular we consider the affect of an expanding universe on shell collapse configurations which may play a significant role in the primordial past and in the asymptotic future.
\newline

The spacetime region exterior, $r>R$, to the collapsing domain wall is described by the Sultana-Dyer metric, in our case, given by \cite{p1}
\begin{eqnarray}
\left(ds^2\right)^+& &=g^+_{\mu\nu}dx_+^{\mu}dx_+^{\nu}=-\left(1-\frac{2M}{r}\right)d\eta^2+\nonumber\\
& &a^2\left(\eta\right)\left[\frac{dr^2}{1-\frac{2M}{r}}+r^2\left(d\theta^2+sin^2\theta\;d\phi^2\right)\right],\nonumber\\
\label{out_met}& &
\end{eqnarray}
where $a\left(\eta\right)$ is the scale factor of the universe and $R=R\left(\eta\right)$ is the equation of the domain wall radius. $\eta$ is the timelike coordinate for the expanding universe.
\newline

For the current universe the scale factor may be set to unity, from which the Schwarzschild metric \cite{p7} follows
\begin{eqnarray}
\left(ds^2\right)^+& &=g^+_{\mu\nu}dx_+^{\mu}dx_+^{\nu}=-\left(1-\frac{2M}{r}\right)dt^2+\nonumber\\
& &\frac{dr^2}{1-\frac{2M}{r}}+r^2\left(d\theta^2+sin^2\theta\;d\phi^2\right).\nonumber\\
\label{schwarz}& &
\end{eqnarray}

To measure the time taken for a black hole to form for an observer at infinity, one may consider the measured time interval of photons travel time between the initial radius, $R_0$, and the final horizon radius, $R_s=2M$, such that they are radially propagating, which classically gives a Schwarzschild time interval of
\begin{equation}
\Delta t=\displaystyle\int_{R_s}^{R_0}\frac{dr}{1-\frac{2M}{r}}\rightarrow\infty.
\label{hor_for_tim_tak}
\end{equation}
This is a non-finite result in this case. Thus such an observer will not measure the creation of the black hole while in the black hole proper frame the horizon and furthermore the black hole itself will exist.
\newline

For the asymptotic observer in a universe with a changing scale factor, the time interval for the production of a black hole from a shell is generalized to
\begin{equation}
\displaystyle\int_{t_f}^{t_i}\frac{d\eta}{a^2\left(\eta\right)}=\displaystyle\int_{R_s}^{R_0}\frac{dr}{1-\frac{R_s}{r}}\rightarrow\infty,
\label{class_res_infty}
\end{equation}
where the time interval will depend on the exact model used and the background metric has been generalized to Eq.(\ref{out_met}). However due to the lack of an explicit expression of $a\left(\eta\right)$, it remains uncertain whether the observer at infinity will or will not measure the production of a horizon in a finite time, specific standard cosmological model phases of the universe are discussed and examined later on.
\newline

On the other hand at some point quantum mechanical effects will become significant due to the exceedingly close surface and final horizon surface. When one considers quantum theory the position of the domain wall has an inherent uncertainty attached to it, in particular the horizon radius takes on the form $R=R_s+\delta R_s$, with an uncertainty of $\delta R_s$ attached to the radius. Thus resulting in a time interval
\begin{eqnarray}
\displaystyle\int_{t_f}^{t_i}\frac{d\eta}{a^2\left(\eta\right)}& &=\displaystyle\int_{R_s+\delta R_s}^{R_0}\frac{dr}{1-\frac{R_s}{r}}\\
& &\sim R_s\ln\frac{R_0-R_s}{\delta R_s}.
\end{eqnarray}

If this was indeed the case, there would be deep and profound implications for astrophysical observations, such that classical arbitrary time processes would be observable in finite time for the standard cosmological forms of $a\left(\eta\right)$.
\newline

However if quantum mechanics does play such a role then the inherent uncertainly it implicates can go in both directions, namely $\pm\delta R_s$. An observational advantage for this process would be that over time quantum effects do at some point, in some instances, propel the realization of the domain wall to such an extent within its classical radius that a black hole does indeed form due to the extreme strength of gravity and its potential to overcome all other known forces in such situations. In other cases this uncertainly maintains the classical result of infinite production time.
\newline

Setting the domain wall model into more concrete terms, we consider a Nambu-Goto spherical domain wall as in Ref.\cite{p8,p12}. The action is given by
\begin{align}
S&=\displaystyle\int d^4x\sqrt{-g}\left[-\frac{1}{16\pi}R+\frac{1}{2}\left(\partial_{\mu}\Phi\right)^2\right]\nonumber\\
&-\sigma\displaystyle\int d^3\zeta\sqrt{-\zeta}+S_{obs},
\label{orig_action}
\end{align}
where the first tern is the Einstein-Hilbert action which leads to general relativity, the second is a scalar field for the action, this may open up the possibility for massless scalar fields which could couple to the gravitational field, the third term accounts for the domain wall that is needed to form the black hole in the first place. This last action component is given in terms of the wall coordinates, $\zeta^a$, which forms a $\left(1+2\right)$ domain wall internal coordinate system, in addition the wall tension, $\sigma$, is also included in this action. Lastly the $S_{obs}$ term denotes the action for the observer.
\newline

In this paper we investigate the collapse of domain walls in a universe where the scale factor is allowed to vary over external time, and explore some of the observational consequences for formation time intervals. Recently there has been much interest in the quantum behavior of shell collapses such as among others Ref.\cite{p8, p9, p10} which explore this phenomenon as measured both by asymptotic observers and infalling observers, we inspect the former case. Other approaches have been explored in \cite{p17, p18, p19, p20, p21, p22}. In Sec. II we derive a mass formula for the domain wall as it collapses and present the formalism in general for treating such structures. While in Sec. III the semi-classical collapse model as measured by an asymptotic observer is explored. The quantum treatment is then given in Sec. IV, where in both cases we derive the radial equation of the domain collapse. Particular scale factor models are considered in Sec. V as well as consequences for the domain wall in the asymptotic future. Sec. VI summarizes results and includes a discussion.
\newline

Units where $G=1=c=\bar{h}$ will be and have been used. Repeated indices are to be summed and note that the signature $\left(-,+,+,+\right)$ is used throughout. Greek indices are taken to refer to the general coordinate system while Latin indices are to refer to hypersurface coordinates as will be shown later on.

\section{II. Domain Wall Mass}
\subsection{A. Experimental Set-up}
\begin{figure}
\centerline{\includegraphics[width=3cm,height=7cm]{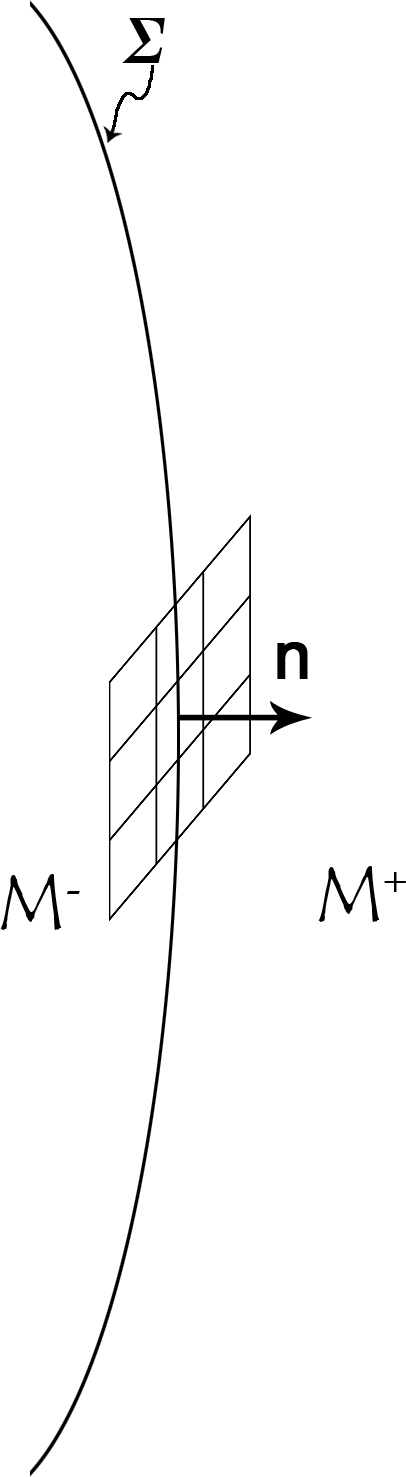}}[h]
\caption{The timelike hypersurface $\Sigma$ is shown which forms the boundary of the individual spacetime regions $M^-$ and $M^+$. The normal \textbf{n} is also depicted.}
\label{hyper}
\end{figure}

Domain walls in a real sense represent jump discontinuities in the stress-energy tensor which in turn means that the background spacetime metric contains the same analogues discontinuity. To investigate this class of problems Israel in Ref.\cite{p2} develops a mathematical framework in which such problems can be differentiated and explored to their natural end, in the classical sense. The particular scenario that is expounded upon in his framework is the case where a spacetime $\mathcal{M}$ is separated into two distinct and discontinuous regions $\mathcal{M}^{\pm}$, with line-elements given by $\left(ds^2\right)^{\pm}=g_{\mu\nu}^{\pm}dx_{\pm}^{\mu}dx_{\pm}^{\nu}$, where the common boundary $\Sigma$ forms a hypersurface with an induced line-element $d\sigma^2=h_{ij}dx^idx^j$. $\mathcal{M}^+$ and $\mathcal{M}^-$ are respectively the exterior and interior spacetime regions to the induced surface, a configuration which is illustrated in Fig.(\ref{hyper}). The common boundary is hence defined by
\begin{equation}
\partial M^+\cap\partial M^-=\Sigma,
\end{equation}
which in a more concise way is why a hypersurface results.

This led to an extensive study of domain walls in general with a focus on mass formulae in terms of wall parameters such as in Refs.\cite{p3, p4, p5, p6}, which we here expand to include cosmological black holes \cite{p1}.
\newline

The unit normal to $\Sigma$ is introduced as the vector \textbf{n} pointing from $\mathcal{M}^-$ to $\mathcal{M}^+$, with the scalar product used in a constraining manner through
\begin{equation}
\textbf{n}\cdot\textbf{n}=g_{\mu\nu}n^{\mu}n^{\nu}\equiv\epsilon=\begin{cases}1,\;\;&\text{if}\;\Sigma\;\text{is timelike}\\-1,\;\;&\text{if}\;\Sigma\;\text{is spacelike.}\end{cases}
\end{equation}

Since the surface $\Sigma$ is formed by the geometry of the two regions $\mathcal{M}^{\pm}$, the extrinsic curvature may be used to compare and contrast the two geometries. In particular the extrinsic curvature on either side $K^{\pm}_{\mu\nu}$ is taken to be the \textbf{n}-component of the covariant derivative in one of the regions $\mathcal{M}^{\pm}$ on a vector $\textbf{e}_{\mu}$ in $\Sigma$, giving
\begin{equation}
K^{\pm}_{\mu\nu}=\textbf{n}\cdot\nabla_{\mu}^{\pm}\textbf{e}_{\nu}=\epsilon n_{\alpha}\Gamma^{\alpha}_{\mu\nu}\mid^{\pm}.
\label{kdef}
\end{equation}

Now for the induced metric to have a non-vanishing surface to map to, the metrics spanning the two regions must agree on the surface in question, however the embedding and so the extrinsic curvature need not agree. Defining the induced metric to be the projection
\begin{equation}
h_{\mu\nu}^{\pm}=g_{\mu\nu}^{\pm}-\epsilon n_{\mu}^{\pm}n_{\nu}^{\pm},
\end{equation}
meaning that there must exist a transformation of some sort on $\Sigma$ that relates $h_{\mu\nu}^+$ with $h_{\mu\nu}^-$, allowing us to set $h_{\mu\nu}^{\pm}=h_{\mu\nu}$ since this is a surface metric only.
\newline

Considering now the Gauss Theorema Egregium and the Codazzi equation in a suitable form \cite{p6, p7} which can be respectively given by
\begin{eqnarray}
^{\left(3\right)}\udt{R}{\alpha}{\beta\mu\nu}&=&\udt{R}{\lambda}{\gamma\sigma\rho}\udt{h}{\alpha}{\lambda}\udt{h}{\gamma}{\beta}\udt{h}{\rho}{\nu}+\nonumber\\
&&\epsilon\left(\udt{K}{\alpha}{\mu}K_{\beta\nu}-\udt{K}{\alpha}{\nu}K_{\beta\mu}\right),\nonumber\\
&&\\
^{\left(3\right)}\nabla_{\alpha}K_{\beta\mu}&-&^{\left(3\right)}\nabla_{\mu}K_{\beta\alpha}=\udt{R}{\lambda}{\sigma\rho\delta}n_{\lambda}\udt{h}{\sigma}{\beta}\udt{h}{\rho}{\alpha}\udt{h}{\delta}{\mu},\nonumber\\
&&
\end{eqnarray}
two relationships result which hold for both exterior and interior regions
\begin{eqnarray}
E_{\mu\nu}n^{\mu}n^{\nu}&=&-\frac{1}{2}\epsilon ^{\left(3\right)}R+\frac{1}{2}\left(K^2-K_{\alpha\beta}K^{\alpha\beta}\right),\nonumber\\
&&\label{Gauss}\\
E_{\mu\nu}\udt{h}{\mu}{\alpha}n^{\nu}&=&-\left(^{\left(3\right)}\nabla_{\mu}\udt{K}{\mu}{\alpha}-^{\left(3\right)}\nabla_{\alpha}K\right).\label{cod}
\end{eqnarray}
Einstein's field equations are above taken to be
\begin{equation}
E^{\pm}_{\mu\nu}=\kappa T^{\pm}_{\mu\nu},
\end{equation}
where $R_{\mu\nu}$ is the Ricci curvature tensor, $R=g^{\mu\nu}R_{\mu\nu}$ is the Ricci scalar, $T_{\mu\nu}$ is the stress-energy tensor, $\kappa=8\pi$ and $E_{\mu\nu}=R_{\mu\nu}-\frac{1}{2}Rg_{\mu\nu}$ is the Einstein tensor.
\newline

Due to the fact of there being two distinct spacetime regions separated by a boundary surface $\Sigma$, we define the following operators to simplify matters later on
\begin{eqnarray}
\left[A\right]&&\equiv A^+-A^-,\label{op1}\\
\{A\}&&\equiv\frac{1}{2}\left(A^++A^-\right).
\end{eqnarray}
Next the Lanczos equation is introduced \cite{p6}
\begin{equation}
\left[K_{ij}\right]-h_{ij}\left[K\right]=\epsilon\kappa S_{ij},
\label{lanc}
\end{equation}
where $K_{ij}$ is the intrinsic curvature and $\epsilon$ is set to $1$ when $\Sigma$ is timelike, and $-1$ when $\Sigma$ is a spacelike surface. Latin indexes are used to refer to the surface metric on the common boundary domain.
\newline

By contracting Eq.(\ref{lanc}) with $h_{ij}$ and substituting the result back into Eq.(\ref{lanc}), the equality
\begin{equation}
\left[K_{ij}\right]=\kappa\epsilon\left(S_{ij}-\frac{1}{2}h_{ij}S\right),
\end{equation}
follows.
\newline

Thus relating the stress-energy tensor with the different $\Sigma$ surfaces that may be considered. $\Sigma$ will be taken to be a timelike surface in the following, and so we let $\epsilon=1$. Also we use the identity \cite{p6}
\begin{equation}
\left[AB\right]=\left[A\right]\{B\}+\{A\}\left[B\right].
\end{equation}

Applying the operator defined in Eq.(\ref{op1}) and taking the Lanczos equation given in Eq.(\ref{lanc}), the relations Eq.(\ref{Gauss}, \ref{cod}) can be reformulated to give
\begin{eqnarray}
^{\left(3\right)}\nabla_j\udt{S}{j}{i}+\left[T_{in}\right]&=&0,\label{cse}\\
S_{ij}\{K^{ij}\}+\left[T_{nn}\right]&=&0,\label{cse2}
\end{eqnarray}
which are what will lead in the next part to a hold on the domain wall mass formula.

\subsection{B. Mass}
Moving now onto the effects of a stress-energy tensor of this sort in a cosmological context. At the jump discontinuity, between the interior and exterior of the domain wall, a surface stress-energy tensor is considered of the general form
\begin{align}
S^{ij}&=\sigma u^iu^j,\nonumber\\
u^iu_i&=-1,
\end{align}
which describes a dust shell and where $\sigma$ is the mass-energy density of the shell. The $u^i$ components of the co-moving velocity will be tangent to the surface layer.
\newline

For the region interior to the domain wall, $r<R$, the spacetime background is described by means of
\begin{eqnarray}
\left(ds^2\right)^-& &=g^-_{\mu\nu}=-dT^2+a^2\left(\eta\right)\Big[dr^2+\nonumber\\
& &r^2\left(d\theta^2+\sin^2\theta d\phi^2\right)\Big],
\label{intr_met}
\end{eqnarray}
which is found by letting the mass vanish in the Sultana-Dyer metric, and where the argument will be suppressed for $a\left(\eta\right)$ in what follows unless otherwise stated. The interior time coordinate, $T$, is related to the asymptotic time, $\eta$, in the exterior by a relation on the shell derived below.
\newline

Considering $\left(1+1\right)-$slices of the interior solution given in Eq.(\ref{intr_met}), one finds the differential
\begin{equation}
\alpha=\frac{dT}{d\tau}=\sqrt{1+a^2 R_{\tau}^2},
\label{alpha_time_diff}
\end{equation}
where $R_{\tau}=\frac{dR}{d\tau}$.
\newline

Next we consider $\left(1+1\right)-$slices of the exterior solution in Eq.(\ref{out_met}) which gives
\begin{equation}
\beta=\frac{d\eta}{d\tau}=\frac{1}{1-\frac{2M}{R}}\sqrt{1-\frac{2M}{R}+a^2 R_{\tau}^2},
\label{tim}
\end{equation}
this finally results in the combination
\begin{equation}
\dot{T}=\frac{dT}{d\eta}=\sqrt{1-\frac{2M}{R}-\frac{\frac{2M}{R}a^2\dot{R}^2}{1-\frac{2M}{R}}},
\end{equation}
where dots represent differentiation with respect to external time $\eta$. The above relation only holds on the shell of the domain wall.

In light of the vanishing stress-energy tensor and due to Eq.(\ref{cse}), it follows that
\begin{eqnarray}
^{\left(3\right)}\nabla_j\udt{S}{j}{i}&=&^{\left(3\right)}\nabla_j\left(\sigma u^iu^j\right)\nonumber\\
&=&u^i\,^{\left(3\right)}\nabla_j\left(\sigma u^j\right)+\sigma u^j\,^{\left(3\right)}\nabla_j u^i\nonumber\\
&=&0.
\end{eqnarray}

Contracting with $u_i$ and utilizing the identity $u_ia^i\equiv u_iu^j\,^{\left(3\right)}\nabla_ju^i=0$, where $a^i$ are the components of the acceleration
\begin{equation}
^{\left(3\right)}\nabla_j\left(\sigma u^j\right)=0,
\end{equation}
which implies a conserved particle number, and furthermore
\begin{equation}
u^j\,^{\left(3\right)}\nabla_ju^i=0,
\end{equation}
meaning that the dust particles are freely falling giving geodesic worldline paths.
\newline
Secondly Eq.(\ref{cse2}) results in
\begin{equation}
S_{ij}\{K^{ij}\}=\sigma u_iu_j\{K^{ij}\}=0.
\end{equation}

The line-element, $h_{ij}$, for the $\left(1+2\right)-$shell is described by
\begin{equation}
ds^2=-d\tau^2+a^2\left(\eta\right)\left[R^2\left(\tau\right)\left(d\theta^2+\sin^2\theta d\phi^2\right)\right],
\end{equation}
where the proper time, $\tau$, is used since the shell is in its proper coordinate system. In the case of the corresponding metric for the domain wall, Eq.(\ref{cse}) results in the proper time derivative
\begin{equation}
\sigma_{\tau}=-\sigma\frac{1}{\sqrt{|h|}}\left(\sqrt{|h|}u^j\right)_{,j},
\end{equation}
where $h$ is the determinant of the $h_{ij}$ and spatial derivatives of $\sigma$ vanish due to the uniform distribution of mass-energy about the shell. Note that
\begin{equation}
h=-a^4\left(\eta\right)R^4\left(\tau\right)\sin^2\theta.
\end{equation}
Thus giving
\begin{equation}
\sigma_{\tau}=-2\sigma\left(\frac{1}{a}a_{\tau}+\frac{R_{\tau}}{R}\right),
\end{equation}
and integrating
\begin{equation}
\sigma=\frac{A}{a^2R^2},
\end{equation}
where $A$ is a constant. Furthermore the shell mass is given by
\begin{equation}
\mu=4\pi R^2\sigma,
\label{res_mas}
\end{equation}
which is clearly a constant because evaporative and back reaction forces are being ignored since they are not important for the aim of this analysis, namely formation times for cosmological black holes.
\newline

Considering now the four-velocity as measured from an observer outside the domain wall, this is given by
\begin{equation}
u_+^{\alpha}=\left(\eta_{\tau},R_{\tau},0,0\right),
\label{vel}
\end{equation}
while by holding to the condition $u^{\alpha}n_{\alpha}|^{\pm}=0$, the vector $n_{\alpha}$ is found to be
\begin{equation}
n_{\alpha}^+=\left(-R_{\tau},\eta_{\tau},0,0\right),
\label{nor}
\end{equation}
and thus placing the restriction
\begin{equation}
u^{\alpha}u_{\alpha}|^+=\eta_{\tau}^2\ddot{g}^++R_{\tau}^2g_{rr}^+=-1.
\end{equation}

Taking the covariant derivative, $u^{\beta}\nabla_{\beta}$, of the velocity identity $u_{\alpha}u^{\alpha}=-1$ and substituting, it is found that
\begin{equation}
n_{\alpha}a^{\alpha}|^+=\left(n_r-\dot{n}\frac{u_r}{\dot{u}}\right)u^{\beta}\nabla_{\beta}u^r|^+.
\label{nacc}
\end{equation}

For the same region all the components of the acceleration vanish except for the radial part given by
\begin{eqnarray}
a^r&=&u^{\beta}\nabla_{\beta}u^r|^{\pm}\nonumber\\
&=&\udt{u}{r}{,\beta}u^{\beta}|^++\Gamma^r_{\alpha\beta}u^{\alpha}u^{\beta}|^+,
\end{eqnarray}
where the second term is given by
\begin{equation}
\Gamma^r_{\alpha\beta}u^{\alpha}u^{\beta}|^+=a^2\left[2aR_{\tau}\dot{a}\sqrt{1-\frac{2M}{R}+a^2R_{\tau}^2}+\frac{M}{R^2}\right].
\end{equation}
Hence giving a final acceleration of
\begin{equation}
a^r=R_{\tau\tau}+a^2\left[2aR_{\tau}\dot{a}\sqrt{1-\frac{2M}{R}+a^2R_{\tau}^2}+\frac{M}{R^2}\right].
\end{equation}
Using Eq.(\ref{vel}), Eq.(\ref{nor}) and Eq.(\ref{tim})
\begin{equation}
\left(n_r-n_{\eta}\frac{u_r}{u_{\eta}}\right)|^+=\frac{1}{\sqrt{1-\frac{2M}{R}+a^2R_{\tau}^2}}.
\end{equation}
Giving Eq.(\ref{nacc}) in terms of domain wall parameters
\begin{eqnarray}
n_{\alpha}a^{\alpha}|^+&=&\frac{1}{\sqrt{1-\frac{2M}{R}+a^2R_{\tau}^2}}\Bigg[\nonumber\\
& &a^2\left(2aR_{\tau}\dot{a}\sqrt{1-\frac{2M}{R}+a^2R_{\tau}^2}+\frac{M}{R^2}\right)\nonumber\\
& &+R_{\tau\tau}\Bigg].
\end{eqnarray}

Setting $M=0$ corresponds to the same quantity for the interior region of the domain wall
\begin{equation}
n_{\alpha}a^{\alpha}|^-=\frac{1}{\sqrt{1+a^2R_{\tau}^2}}\left[2a^3R_{\tau}a_T\sqrt{1+a^2R_{\tau}^2}+R_{\tau\tau}\right].
\end{equation}

Contracting the geodesic equation, $u^{\alpha}\nabla_{\alpha}u^{\beta}|^{\pm}$, and employing Eq.(\ref{kdef}), the orthogonal components of the acceleration are found to be
\begin{equation}
n_{\alpha}a^{\alpha}|^{\pm}=K^{\pm}_{ij}u^iu^j.
\label{mid}
\end{equation}

Along with this and the vanishing stress-energy form of Eq.(\ref{cse2}), the following pivotal relationship emerges
\begin{equation}
n_{\alpha}a^{\alpha}|^++n_{\alpha}a^{\alpha}|^-=0.
\label{add}
\end{equation}

Now by considering the Lanczos equation in Eq.(\ref{lanc}) and Eq.(\ref{mid}) for this stress-energy tensor
\begin{equation}
\left[a^{\alpha}n_{\alpha}\right]=\frac{\kappa}{2}\sigma,
\end{equation}
which when coupled with Eq.(\ref{add}) it follows that
\begin{equation}
-2a^{\alpha}n_{\alpha}|^-=\frac{\kappa}{2}\sigma=4\pi\sigma.
\label{final}
\end{equation}

Also by Eq.(\ref{add}), the acceleration of the domain wall is found in terms of lesser derivatives to be
\begin{eqnarray}
R_{\tau\tau}&=&\left(-2a^3R_{\tau}\left[\frac{da}{dT}+\dot{a}\right]-\frac{a^2\frac{M}{R^2}}{\sqrt{1-\frac{2M}{R}+a^2R_{\tau}}}\right)\nonumber\\
& &\frac{\sqrt{1-\frac{2M}{R}+a^2R_{\tau}^2}\sqrt{1+a^2R_{\tau}^2}}{\sqrt{1-\frac{2M}{R}+a^2R_{\tau}^2}+\sqrt{1+a^2R_{\tau}^2}},
\end{eqnarray}
which substituted into Eq.(\ref{final}) gives the mass of the domain wall
\begin{equation}
M=\frac{\frac{1}{2}\left[\sqrt{1-\frac{2M}{R}+a^2R_{\tau}^2}+\sqrt{1+a^2R_{\tau}^2}\right]4\pi R^2\sigma}{a^2-\frac{4a^3}{R^3}R_{\tau}\dot{a}\sqrt{1-\frac{2M}{R}+a^2R_{\tau}^2}}.
\label{mass_for}
\end{equation}
This reduces to the mass function found in Ref.\cite{p3} when the scale factor is set to unity, as was expected. The generalization is important because it incorporates the changing scale factor of the universe which has effects in some, cosmological, cases, and will certainly have effects on black hole formation due to the long time scale on which such black holes are expected to take to form.

For relatively fast cosmological events as they in general are or for large event horizons, the second term in the denominator can be ignored, and thus Eq.(\ref{mass_for}) becomes
\begin{align}
M\approx&\frac{1}{2a^2}\left[\sqrt{1-\frac{2M}{R}+a^2R_{\tau}^2}+\sqrt{1+a^2R_{\tau}^2}\right]\nonumber\\
&4\pi R^2\sigma,
\label{mass_cons_law}
\end{align}
which is still heavily dependent on the particular scale factor function $a\left(\eta\right)$. Next solving for $M$ explicitly in terms of the proper radial velocity
\begin{equation}
M=\frac{4\pi R^2\sigma}{a^4}\left[a^2\sqrt{1+a^2R_{\tau}^2}-2\pi R\sigma\right],
\label{mas_for_ham}
\end{equation}
which in interior coordinate time, $T$, turns out to be given by
\begin{equation}
M=\frac{4\pi\sigma R^2}{a^4}\left[\frac{a^2}{\sqrt{1-a^2R^2_{T}}}-2\pi\sigma R\right],
\label{mass_ham_sphe_anz}
\end{equation}
due to a relationship that emerges through Eq.(\ref{alpha_time_diff}), namely
\begin{equation}
\sqrt{1+a^2\dot{R}^2}=\frac{1}{\sqrt{1-a^2R^2_T}}.
\end{equation}

Considering Eq.(\ref{mas_for_ham}) and its physical interpretation, the first term in the numerator represents the total rest mass of the shell, whereas the second term accounts for the binding energy or self-gravity of the domain wall, hence this forms the total energy of the domain wall, and so can be considered as the Hamiltonian $H\equiv M$.

\section{III. Semi-Classical Observer at Infinity Treatment}
Placing the spherical ansatz in Eq.(\ref{mass_ham_sphe_anz}) along with the metric into the original action in Eq.(\ref{orig_action}) leads to an effective action for the radial coordinate. However this does not lead to the correct dynamics for the gravitating system, and moreover does not lead to the mass conservation law in Eq.(\ref{mass_cons_law}). Thus an alternative approach \cite{p8} must be taken, namely that of taking an appropriate ansatz for the Lagrangian followed by comparison with the Hamiltonian through the Hamiltonian Legendre transformation. Giving the correct effective action
\begin{equation}
S_{eff}=-\frac{4\pi\sigma}{a^4}\displaystyle\int dT R^2\left[a^2\sqrt{1-a^2R^2_T}-2\pi\sigma R\right],
\end{equation}
which when written in terms of external time becomes
\begin{align}
S_{eff}&=-\frac{4\pi\sigma}{a^4}\displaystyle\int d\eta R^2\Bigg[a^2\sqrt{1-\frac{2M}{R}-\frac{a^2\dot{R}^2}{1-\frac{2M}{R}}}\nonumber\\
&-2\pi\sigma R\sqrt{1-\frac{2M}{R}-a^2\dot{R}^2\frac{2M/R}{1-\frac{2M}{R}}}\Bigg].
\end{align}
Thus the associated Lagrangian turns out to be
\begin{align}
L_{eff}&=-\frac{4\pi\sigma}{a^4} R^2\Bigg[a^2\sqrt{1-\frac{2M}{R}-\frac{a^2\dot{R}^2}{1-\frac{2M}{R}}}\nonumber\\
&-2\pi\sigma R\sqrt{1-\frac{2M}{R}-a^2\dot{R}^2\frac{2M/R}{1-\frac{2M}{R}}}\Bigg].
\end{align}
Finding the radial velocity derivative yields the generalized momentum, $\Pi$, such that
\begin{align}
\Pi&=\frac{\partial L_{eff}}{\partial\dot{R}}=\frac{4\pi\sigma R^2\dot{R}}{a^2\sqrt{1-\frac{2M}{R}}}\Bigg[\frac{a^2}{\sqrt{\left(1-\frac{2M}{R}\right)^2-a^2\dot{R}^2}}\nonumber\\
&-\frac{4\pi\sigma M}{\sqrt{\left(1-\frac{2M}{R}\right)^2-\frac{2Ma^2\dot{R}^2}{R}}}\Bigg],
\label{gen_mom_exp}
\end{align}
again employing the Hamiltonian Legendre transformation (in terms of $\dot{R}$) yields a Hamiltonian
\begin{align}
H&=\frac{4\pi\sigma R^2}{a^2}\left(1-\frac{2M}{R}\right)^{3/2}\Bigg[\frac{1}{\sqrt{\left(1-\frac{2M}{R}\right)^2-a^2\dot{R}^2}}\nonumber\\
&-\frac{1}{a^2}\frac{2\pi\sigma R}{\sqrt{\left(1-\frac{2M}{R}\right)^2-\frac{2M}{R}a^2\dot{R}^2}}\Bigg].
\end{align}
Now to eliminate $\dot{R}$ in favor of $\Pi$ would mean using Eq.(\ref{gen_mom_exp}) and solving a quartic. Alternatively we consider the near field of the horizon, as $R$ approaches $2M$, when the denominators are approximately equal and the generalized momentum approaches
\begin{equation}
\Pi\approx\frac{4\pi\mu R^2\dot{R}}{a^2\sqrt{1-\frac{2M}{R}}}\frac{1}{\sqrt{\left(1-\frac{2M}{R}\right)^2-a^2\dot{R}^2}},
\end{equation}
Taking
\begin{equation}
\mu\left(\eta\right)\equiv\sigma\left(a^2-4\pi\sigma M\right),
\end{equation}
resulting in a Hamiltonian equivalent to
\begin{align}
H&\approx\frac{4\pi\mu R^2\left(1-\frac{2M}{R}\right)^{3/2}}{a^4\sqrt{\left(1-\frac{2M}{R}\right)^2-a^2\dot{R}^2}}\nonumber\\
&=\Bigg[\left(\left(1-\frac{2M}{R}\right)a^3\Pi\right)^2\nonumber\\
&+\left(1-\frac{2M}{R}\right)\left(4\pi\mu R^2\right)^2\Bigg]^{1/2},
\label{approx_hamil_simp}
\end{align}
and so the Hamiltonian has the form of the energy of a relativistic particle, $\sqrt{p^2+m^2}$, with a position dependent mass measurement.
\newline

Considering next the fact that the Hamiltonian is a conserved quantity, it may be taken as a constant of motion such that
\begin{equation}
\frac{\mu R^2\left(1-\frac{2M}{R}\right)^{3/2}}{\sqrt{\left(1-\frac{2M}{R}\right)^2-a^2\dot{R}^2}}=h,
\label{const_hamil_approx}
\end{equation}
where $\text{h}=\text{H}/4\pi$ is a constant (up to the approximation used in obtaining the simpler form of the Hamiltonian in Eq.(\ref{approx_hamil_simp})).
\newline

Solving Eq.(\ref{const_hamil_approx}) for $\dot{R}$
\begin{equation}
\dot{R}=\pm\frac{1-\frac{2M}{R}}{a}\sqrt{1-\left(1-\frac{2M}{R}\right)\frac{\mu^2 R^4}{h^2}},
\end{equation}
which near the horizon takes the form
\begin{equation}
\dot{R}\approx\pm\frac{1-\frac{2M}{R}}{a}\left(1-\frac{1}{2}\left(1-\frac{2M}{R}\right)\frac{\mu^2 R^4}{h^2}\right).
\end{equation}

The dynamics for $R\sim\dot{R}$ can thus be obtained by solving the equation $\dot{R}=\pm\frac{\left(1-\frac{2M}{R}\right)}{a}$, which to leading order in $R-2M$ yields,
\begin{equation}
R\left(\eta\right)\approx 2M+\left(R_0-2M\right)e^{\pm\frac{1}{2M}\displaystyle\int\frac{d\eta}{a\left(\eta\right)}},
\label{rad_vel}
\end{equation}
where $R_0$ is the radius of the domain wall when $\eta=0$.
\newline

The negative sign is chosen since collapsing solutions are being investigated but the positive sign would describe expanding domain walls such as simple big bang investigations. The solution above implies that the asymptotic classical observer does not ever measure the formation of the black hole horizon since $R\left(\eta\right)=2M$ only when $\displaystyle\int\frac{d\eta}{a\left(\eta\right)}\rightarrow\infty$, which is as expected since when $a\left(\eta\right)=1$, the asymptotic time taken for the domain to reach the future horizon grows as in Eq.(\ref{hor_for_tim_tak}), a fact which should not change with the size of the universe, as long as its size is not smaller than the future horizon itself. However in that case, since the black hole would have already formed in a real sense, and thus existed throughout the process, the foundational supposition of the above analysis would have been broken, that the black hole must not have pre-existed and so the background metric within the domain wall can be described by a flat metric given in Eq.(\ref{out_met}). Quantum effects are now probed for changes in this conclusion.

\section{IV. Quantum Treatment}
Taking the square of the classical Hamiltonian
\begin{align}
H^2&=\left(1-\frac{2M}{R}\right)a^3\Pi\left(1-\frac{2M}{R}\right)a^3\Pi\nonumber\\
&+\left(1-\frac{2M}{R}\right)\left(4\pi\mu R^2\right)^2,
\label{sq_ham_quan}
\end{align}
where a choice has been made as regards the ordering of $\left(1-\frac{2M}{R}\right)a^3$ and $\Pi$. This quantity will be used later in the Schrödinger equation of the domain wall. In general, terms are added that depend on the commutator $\left[\left(1-\frac{2M}{R}\right)a^3,\Pi\right]$ as throughout modern quantum mechanics. In the limit $R\rightarrow2M$
\begin{equation}
\left[\left(1-\frac{2M}{R}\right),\Pi\right]\sim\frac{1}{2Ma^3}.
\end{equation}

Taking $H$ to be the mass, $M$, of the domain wall, the terms due to the operator order ambiguity will be negligible provided that
\begin{equation}
M\gg\frac{1}{2Ma^3}\sim\frac{M_p}{Ma^3},
\end{equation}
where $M_p$ is the Planck mass. Hence the order ambiguity for domain walls much greater than the Planck mass can be ignored.
\newline

Applying now the standard quantization procedure, we let
\begin{equation}
\Pi=-i\frac{\partial}{\partial R},
\end{equation}
and considering the squared time-dependent Schrödinger equation
\begin{equation}
H^2\Psi=-\frac{\partial^2\Psi}{\partial\eta^2},
\end{equation}
yields
\begin{align}
&-a^6\left(1-\frac{2M}{R}\right)\frac{\partial}{\partial R}\left(\left(1-\frac{2M}{R}\right)\frac{\partial\Psi}{\partial R}\right)\nonumber\\
&+\left(1-\frac{2M}{R}\right)\left(4\pi\mu R^2\right)^2\Psi=-\frac{\partial^2\Psi}{\partial\eta^2}.
\end{align}
In order to solve this equation, we let
\begin{equation}
u=\frac{1}{a^3}\left(R+2M\ln\left|\frac{R}{2M}-1\right|\right),
\label{ur_trans}
\end{equation}
which along with
\begin{equation}
a^3\left(1-\frac{2M}{R}\right)\Pi=-i\frac{\partial}{\partial u},
\end{equation}
is used to convert the squared Schrödinger equation into
\begin{equation}
\frac{\partial^2\Psi}{\partial\eta^2}-\frac{\partial^2\Psi}{\partial u^2}+\left(1-\frac{2M}{R}\right)\left(4\pi\mu R^2\right)^2\Psi=0.
\label{wave_eqn_coll_dom}
\end{equation}

This is just the massive wave equation in a Minkowski background with a mass measurement that depends on position. Note that $R$ can be written in terms of $u$, in principle, by means of the transformation in Eq.(\ref{ur_trans}), taking care of course to choose the correct branch in turn, since $R\in\left(2M,+\infty\right)$ transforms to $u\in\left(-\infty,+\infty\right)$, and $R\in\left(0,2M\right)$ similarly corresponds to $u\in\left(-\infty,0\right)$.
\newline

In order to investigate the situation of the collapsing domain wall, the region $R\sim2M$ is examined, noting that the logarithm in $u\left(R\right)$ dominates in this region, so that
\begin{equation}
R=2M+2Me^{\frac{a^3u}{2M}},
\label{rad_quan_for}
\end{equation}
giving wave-packet solution propagating toward the $R=2M$ surface since as $u\rightarrow-\infty$,
\begin{equation}
\left(1-\frac{2M}{R}\right)\sim e^{\frac{a^3u}{2M}}\rightarrow0,
\end{equation}
and where finally the last term in the massive wave equation can be ignored.
\newline

Wave packet dynamics in this region are simply given by the free wave equation and any such function of the light-cone coordinates $\left(u\pm\eta\right)$ is a solution. In particular taking a Gaussian wave packet solution that is propagating toward the horizon radius
\begin{equation}
\Psi=\frac{1}{\sqrt{2\pi}S}e^{-\left(u+t\right)^2/2s^2},
\end{equation}
where $s$ is some chosen width of the wave packet in the $u$ coordinate. The width of the Gaussian wave-packet remains fixed in the $u$ coordinate while it shrinks in the $R$ coordinate through the relation $dR=a^3\left(1-\frac{2M}{R}\right)du$ which follows by the transformation. This fact reinforces the classical result that even the horizon is not seen to ever form by an outside observer, since if the wave packet were to remain constant in size in the $R$ coordinate, then it might cross the horizon in some finite time, however this is not the case.
\newline

The wave equation in Eq.(\ref{wave_eqn_coll_dom}) implies that in the $u$ coordinate the wave packet travels at the speed of light, however since the horizon is located at $u\rightarrow-\infty$ it will still take an arbitrary large amount of time for the domain wall to collapse to the future horizon radius. Hence the classical conclusion for the asymptotic observer stands inspite of the quantum treatment.
\newline

The massive wave equation was considered using the squared classical Hamiltonian with the inherent order ambiguity of the term, which was necessary to eliminate the square root from occurring. Other quantization procedures may result in a different final conclusions, for example instead of following this procedure we could have taken the near horizon approximation
\begin{align}
H&=\Bigg[\left(\left(1-\frac{2M}{R}\right)a^3\Pi\right)^2\nonumber\\
&+\left(1-\frac{2M}{R}\right)\left(4\pi\mu R^2\right)^2\Bigg]^{1/2}\nonumber\\
&\approx\pm\left(\left(1-\frac{2M}{R}\right)a^3\Pi\right)^2,
\end{align}
where signs are chosen so as to render a non-negative Hamiltonian. The same result follows in this particular case, namely the time-dependent Schrödinger equation yields wave packets propagating at the speed of light in the $u$ coordinate and with a horizon located at $u\rightarrow-\infty$.

\section{V. The Scale Factor}
Considering now the specific theory of the universe offered by the standard cosmological model and the FLRW-metric in particular, the scale factor assumes three distinct forms dependent on the phase that the universe is in. For the primordial universe, the equations of state of the mass within the universe can be approximated by the gas equation of state due to the extreme conditions that existed, such as the hot and denseness of these matter fields, known as the radiation-dominated era, an associated scale factor of
\begin{equation}
a\left(\eta\right)\propto\eta^{1/2},
\end{equation}
is attached. The universe eventually cooled down enough to allow matter to form thus giving birth to the matter-dominated era. This era has the defining feature of primordial matter structures, which approximated by a pressureless gas gives a scale factor of
\begin{equation}
a\left(\eta\right)\propto\eta^{2/3}.
\end{equation}
Finally the present age was realized, namely the dark-energy dominated era, where the equation of state assumes an entirely different form, that of vacuum energy leading to an exponential expansion of the universe such that the scale factor becomes
\begin{equation}
a\left(\eta\right)\propto e^{\sqrt{\frac{\Lambda}{3}}\eta}.
\end{equation}

The mass of the domain wall given in Eq.(\ref{mass_for}) in all of the above cases results in a scenario where the mass function in the eternal future vanishes as measured by the asymptotic observer. This is to be expected since as the acceleration of the universe continues to increase mass structures are expected to either form very dense compact structures such as black holes or in turn be ripped apart in the asymptotic future.
\newline

As for the radial component of the domain wall, classically this is given by Eq.(\ref{rad_vel}), which reduce to the individual radii
\begin{align}
R_{Rad\_Dom}\left(\eta\right)&\approx2M+\left(R_0-2M\right)e^{-\eta^{1/2}/M},\\
R_{Mat\_Dom}\left(\eta\right)&\approx2M+\left(R_0-2M\right)e^{-\frac{3\eta^{1/3}}{2M}},\\
R_{Mat\_Dom}\left(\eta\right)&\approx2M+\left(R_0-2M\right)e^{-\frac{1}{2M}e^{-\eta}},
\end{align}
for the respective FLRW phases, all resulting in collapse for the domain wall to the expected value of $2M$ in arbitrarily large asymptotic times. The same final result is achieved for the quantum treatment, found through Eq.(\ref{rad_quan_for})
\begin{align}
R_{Rad\_Dom}\left(\eta\right)&\approx2M+2Me^{\frac{\eta^{3/2} u}{2M}},\\
R_{Mat\_Dom}\left(\eta\right)&\approx2M+2Me^{\frac{\eta^2u}{2M}},\\
R_{Mat\_Dom}\left(\eta\right)&\approx2M+2Me^{\frac{e^{3\eta}u}{2M}}.
\end{align}
As shown all confirm the same result that the black hole is only observed to form after an infinite amount of time has past in the external time $\eta$ coordinate.

\section{VI. Conclusion}
In this paper we investigated the behavior of a collapsing domain wall using the Schrödinger equation in an expanding universe characterized by a scale factor $a\left(\eta\right)$, where $\eta$ is the time as measured by an asymptotic observer. We have expanded on the black hole cases looked at in Refs.\cite{p5, p8, p9, p10} in that the scale factor and so certain cosmological models can now be taken into account in calculating the ability of black holes to form as well as the likelihood that a domain mass configuration will collapse into a black hole structure.
\newline

The classical result was shown in Eq.(\ref{class_res_infty}) which yields infinite formations times, followed by a discussion of some quantum ramifications. Next the situation was given some mathematical structure ending in the mass formula for this, a black hole which is just a cosmological version of the Schwarzschild black hole.
\newline

Following that introduction, the collapse of a gravitating spherical cosmological domain wall in both the classical and quantum cases was delved into, ignoring back reaction and evaporative processes. It is sometimes said that quantum mechanical processes may lead to a contradiction of the classical result, however our quantum treatment reinforces that result and hence it follows that this must be a reasonably general result \cite{p8} even in the FLRW cosmological case. This last point was investigated in the last section which revealed that in an FLRW universe domain walls collapse even when it is the case that its expansion is accelerating, that is an arbitrarily large amount of time is still required.
\newline

The mass of a collapsing body will inevitably decrease from its initial value due to evaporation and back reaction, and so the domain wall begins a process of pursuit of the decreasing horizon during the collapse phase. It is only after the associated horizon radius has been surpassed that the black hole can form. Thus the question is not only will the domain wall collapse to form a horizon but also will there be any mass left in the shell by such time as this is realized, that is will the potential horizon vanish, reduce to $R=0$, before the shell producing it reaches its decreasing event horizon radius? It seems unlikely since such processes become weaker as the mass decreases, however further analysis would have to be done to have a definite answer to this question. Furthermore if such an eventuality were to occur then an infalling observer would never cross the event horizon due to its non-existence and so the same observations will be made by this and the asymptotic observer.
\newline

Realistically however such a collapse model must be further complicated by external forces such as external gravitational fields and nearby matter which may disrupt the process and prompt the production of a black hole in finite time irrespective of the natural arbitrary time needed, even during accelerated cosmological expansion.
\newline

The resulting equations produced in this paper are important in that they implant some of the complications necessary for a more realistic model of gravitational collapse, in this case the expansion of the universe was accounted for with a number of particular periods of expansion examined. In any observation of black holes we will only ever be asymptotic observers, in the first hand measurement sense, and so it is only the measurements in this respect that matter irrespective of whether they are astrophysical observations of laboratory experiments. Furthermore it is clearly the case that, with the exception of Planck scale effects, the period of collapse will in the main produce the most energetic and observationally significant phenomena seen by outside observers \cite{p16} and so it is a formation that could yield important answers for observational problems.

\section{Acknowledgments}
The authors wish to thank the Physics Department at the University of Malta for hospitality during the completion of this work. We would also like to thank Joseph Sultana for helpful conversations.

\end{document}